\documentclass[%
 reprint, 
superscriptaddress,
 amsmath,amssymb,
 aps,
hyperref
]{revtex4-2}
\usepackage{soul}
\usepackage{float} 
\usepackage{graphicx}
\usepackage{dcolumn}
\usepackage{bm}
\usepackage[colorlinks,citecolor=blue,urlcolor=blue,linkcolor=blue]{hyperref}
\usepackage{amsmath,amssymb}
\usepackage[usenames]{color}
\usepackage{colortbl}
\usepackage{eso-pic}
\newcommand{\reportnum}[2]{
  \AddToShipoutPictureBG*{%
    \AtPageUpperLeft{%
      \hspace{0.8\paperwidth}%
      \raisebox{#1\baselineskip}{%
        \makebox[0pt][l]{\textnormal{#2}}
  }}}%
}
\begin{document}
\reportnum{-6}{INR-TH-2024-025}

\title{
Can dark-matter Q-balls grow to the mass gap masses?}

\author{Alexander Libanov}%
\email[\bf{e-mail:  }]{libanov.am18@physics.msu.ru}
\affiliation{Institute for Nuclear Research of the Russian Academy of Sciences, 60th October Anniversary Prospect 7a, Moscow 117312, Russia}

\date{\today}
\begin{abstract}
Within the framework of general relativity, it can be shown that gravitational waves are radiated with the merger of massive compact objects. Such gravitational wave signals are observed on Earth on various detectors, in particular, on Laser Interferometer Gravitational Wave Observatory (LIGO) and Virgo. During the operation of these detectors, many events have been detected. Those events are associated with the merger of massive compact objects, however, the nature of some merging objects has not yet been reliably established. This work considers nontopological solitons of dark matter -- Q-balls,
as candidates for the role of massive compact objects. In this work one of the simplest models of Q-balls, the mechanism of their birth during a phase transition in the early Universe and the mechanism of their mass gaining during the evolution of the Universe, which is based on their mutual merger, are considered. As a result, it is analyzed whether Q-balls of dark matter can grow to the mass gap masses and be candidates for the role of massive compact objects.
\end{abstract}
\maketitle
\section{INTRODUCTION}
\label{intro}
It follows from the general theory of relativity that gravitational waves can be radiated during the merger of massive compact objects \cite{Le_Tiec_2017, LIGOScientific:2020iuh, LIGOScientific:2020ufj, LIGOScientific:2020zkf, LIGOScientific:2020kqk, Baxter:2021swn}. Such objects are usually claimed by neutron stars or black holes \cite{buonanno2015sourcesgravitationalwavestheory}. Nevertheless, with the more and more new signals, unusual data began to arise \cite{Krolak:2021cie, Beradze_2021, LIGOScientific:2020zkf, LIGOScientific:2020ufj, LIGOScientific:2020iuh, Woosley:2021xba}.

On August 14th, 2019, at 21:10:39 UTC, GW190814  was detected using LIGO and Virgo. GW190814 is a gravitational wave event received from the merger of two massive compact objects with masses of $22.3 - 24.3$ M$_{\odot}$ and $2.50-2.67$ M$_{\odot}$, respectively \cite{LIGOScientific:2020zkf}. This event indicates that the smaller object is located at the mass boundary between a light black hole and a massive neutron star \cite{Wystub:2021qrn}. It is noteworthy that no electromagnetic signal was received from this event \cite{Abbott_2020}. There are still discussions about the nature of the second object.
One possible scenario is a merger of a black hole and a neutron star. In this case, strong restrictions are imposed on the radius of the neutron star. The other possible scenario is a merger of two black holes \cite{Beradze_2021}, \cite{M_ller_1996}.  However, there are models that suggest that the smaller object is exotic: a gravastar, a quark star, or a boson star \cite{Mazur_2004, Bombaci_2021, Cao_2022, PhysRev.172.1331}.

On January 5, 2020, at 16:24:26 UTC and on January 15, 2020 at 4:23:10 UTC, signals GW200105 and GW200115 were received using LIGO and Virgo, respectively. GW200105 received as a result of rotation and merger of a binary system, the masses of objects in which are  $8.9^{+1.2}_{-1.5}$ M$_{\odot}$ and $1.9^{+0.3}_{-0.2}$ M$_{\odot}$, and  GW200115 is a result of rotation and merger of a binary system, the masses of objects in which are $5.7^{+1.8}_{-2.1}$ M$_{\odot}$ and $1.5^{+0.7}_{-0.3}$ M$_{\odot}$. The masses of the smaller objects are very different from the mass of the second component of the system, from which the GW190814 signal was received. Moreover, these events were again not accompanied by an electromagnetic signal. Hypothetically, these objects could be dark matter stars \cite{Wystub:2021qrn}, \cite{Abbott_2021}.

The astrophysical
understanding of stellar evolution suggests that stellar-mass black holes may not have masses less than $\sim 5 \mbox{ M}_{\odot}$, and neutron stars are expected to
have a maximum mass of $\sim 3 \mbox{ M}_{\odot}$ \cite{Bailyn_1998, _zel_2010, Farr_2011, _zel_2012, Kiziltan_2013, antoniadis2016millisecondpulsarmassdistribution}. It is also worth noting that LIGO and Virgo distinguish objects primarily by their masses that is why the nature of merging objects is difficult to determine only by the signature of gravitational waves. These lead to the fact that there is a mass gap between heavy neutron stars and light black holes, equal to $\sim 3-5 \mbox{ M}_{\odot}$ \cite{Krolak:2021cie}, \cite{Woosley:2021xba}. Thus, the problem arises of explaining some events like GW190814, GW200105 and GW200115, since some objects in such systems that generate gravitational wave signals lie in the mass gap for black holes or violate the equation of state of neutron stars. Moreover, there is still a problem of dark matter in cosmology, since there are a sufficient number of candidates that have not yet been experimentally confirmed \cite{1970ApJ159379R, Bertone:2018krk, Freese_2009}.

Some works suggest that dark matter stars may be components of binary systems of compact objects \cite{Lee_2021, Di_Giovanni_2022, Pacilio_2020}. According to this possible solution to the problems of unusual gravitational wave signals, mass gap and cold dark matter is considered in this work. The model of nontopological solitons -- Q-balls of dark matter -- may be a candidate for the components of such binary systems. Firstly, using this model, one can try to explain the unusual gravitational wave signals received by LIGO and Virgo, without modifying the equation of state of neutron stars. Secondly, Q-balls may be candidates for the role of cold dark matter \cite{Kusenko:2001vu}. Thus, the work considers the parameters of Q-balls, their mass and radius, the mechanism of formation of cosmological Q-balls, and proposes simple mechanisms for increasing Q-ball mass by merging with other Q-balls.

\section{PHYSICS OF Q-BALLS}
\label{PHYSICS OF Q-BALLS}
\subsection{Q-ball parameters}
\label{Q-ball parameters}
The Q-ball is a nontopological soliton.
There are many different models that allow the existence of Q-balls \cite{q-balls}, \cite{Coleman:1985ki}.
This work considers one of the simplest models of Q-balls -- the Friedberg-Lee-Sirlin theory. This theory consists of two scalar fields, one of which is real and the other is complex, and the reason for the existence and stability of the Q-ball is the presence of a charge in the Q-ball. In the Friedberg-Lee-Sirlin theory, it is assumed that particles of the $\chi$ field acquire mass during interaction with some additional scalar field $\varphi$ \cite{PhysRevD.13.2739, FRIEDBERG197632, Krylov:2013qe}. The Lagrangian of Friedberg-Lee-Sirlin theory is
\begin{equation}
    \label{FLSLagrangian}
    \mathcal{L} = \frac{1}{2}(\partial _{\mu} \varphi)^{2}-U(\varphi)+(\partial _{\mu}\chi)^{*}\partial_{\mu}\chi - k^{2}\varphi^{2}\chi^{*}\chi,
\end{equation}
\[
 U(\varphi)=(\varphi^{2}-v^{2})^{2},
\]
where $\chi$ is a complex scalar field, $\varphi$ is real scalar field, $k= h/\lambda^{1/4}$ in notations of Ref. \cite{Krylov:2013qe} and $v$ are some constants \cite{PhysRevD.13.2739}.
The mass of the field $\chi$ in a vacuum is
\begin{equation}
    \label{masschi}
    m_{\chi}=k v,
\end{equation}
This theory has $U(1)$-symmetry,
\[
\chi \to e^{i\alpha}\chi
\]
and, therefore, by the Noether's theorem, there exists a conserved Noether's charge $Q$. The state with the lowest energy with a sufficiently large $Q$ is a spherical Q-ball with conditions,

\[
\begin{cases}
\varphi = 0, \: x<R,
\\
\varphi = v, \: x>R,
\end{cases}
\]
where $R$ is the radius of the Q-ball. 

In the approximation of a sufficiently large $Q$, the radius $R$ and the energy $E$ of the Q-ball are determined by the balance of the energy of the massless field $\chi$ located in the potential well and the potential energy of the massive field $\varphi$ inside. Then the energy is
\begin{equation}
    \label{QballEnergy}
    E(R)=\frac{\pi Q}{R} + \frac{4 \pi}{3}R^{3} U_{0},
\end{equation}
where
\[
U_{0}=U(0)-U(v)= v^4.
\]

By differentiating (\ref{QballEnergy}) by $R$ and equating the derivative to zero, which is the condition for the minimum energy, an expression for the radius of the Q-ball is obtained

\begin{equation}
    \label{QballRadius}
    R_{Q}=\left(\frac{Q}{4}\right)^{1/4}\frac{1}{v}.
\end{equation}

Now, if (\ref{QballRadius}) is substituted into ({\ref{QballEnergy}}), it can be obtained an expression for the mass of the Q-ball

\begin{equation}
\label{QballMass}
    m_{Q} = \frac{4\sqrt{2}\pi}{3}vQ^{3/4},
\end{equation}
assuming that the mass of the Q-ball corresponds to the minimum energy of the Q-ball \cite{PhysRevD.13.2739}, \cite{Krylov:2013qe}.

It is worth noting that there is a stability condition for the Q-ball, which gives a limit on the minimum charge:
\begin{equation}
    \label{Qmin}
    Q_{min} = \frac{m_{Q}}{m_{\chi}}.
\end{equation}

\subsection{First order phase transition}
\begin{figure}[htb]
    \centering
    \includegraphics[width=\columnwidth]{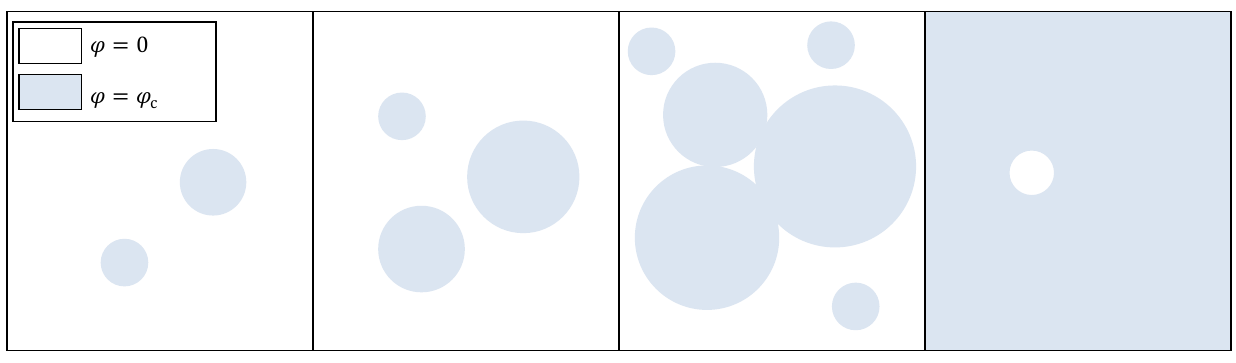}
    \caption{A schematic representation of the I-st order phase transition in the early Universe. The white area is the area of the old phase ($\varphi = 0$), the blue area is the area of the new phase ($\varphi = \varphi_{c}$) \cite{Krylov:2013qe}. As can be seen from the figure, at some point in time, one area of the old phase remains in the allocated volume, which, for simplicity, is considered spherical within the framework of this work.}
    \label{combuction}
\end{figure}

It is assumed that the early expanding flat Universe is uniformly filled with particles of the fields $\varphi$ and $\chi$. The cosmological Q-balls in this work are Q-balls born as a result of the first-order phase transition that occurs due to the cooling of the Universe to some critical temperature $T_{c}$, with $T_{c}\sim v$. During the phase transition, bubbles of a new phase are formed ($\varphi = \varphi_{c}$), which, in turn, merge. Thus, in some volume there remains one bubble of the old phase ($\varphi = 0$), which collects particles of the $\chi$ field under the assumption of sufficient massiveness of the particles of the $\chi$ field, and which is the future cosmological Q-ball (Fig. \ref{combuction}).
This phase transition model does not impose special restrictions on the area of appearance of new phase bubbles during the phase transition, so, theoretically, new phase bubbles can appear in the remaining bubbles of the old phase; therefore, it is assumed that a cosmological Q-ball is formed if and only if a new phase bubble does not form in the  remaining bubble of the old phase \cite{PhysRevD.23.876}, \cite{rubakov2011introduction}.

The physics of the phase transition imposes a restriction on the charge of the cosmological Q-ball from above. In the framework of the work, dark matter is represented by particles of the $\chi$ field, and there is an asymmetry \cite{Krylov:2013qe}
\begin{equation}
    \label{asymmetry}
    \frac{n_{\chi} - n_{\bar{\chi}}}{s} = \frac{n_{Q}Q}{s} = \eta_{\chi},
\end{equation}
where $n_{\chi}$ is the concentration of particles of the field $\chi$, $n_{\bar{\chi}}$ is the concentration of antiparticles of the field $\chi$, $n_{Q}$ is the concentration of Q-balls in the Universe, $s$ is the density of entropy, and $\eta_{\chi} \leq 1$ is the parameter responsible for the asymmetry of the particles of the field $\chi$. This parameter is determined by the dynamics at earlier stages in interactions with other fields, but in the realities of this work it is considered free \cite{Stuke_2011}, \cite{Zarembo_2000}. 

It is possible to estimate the volume of the remaining bubble of old phase, and hence estimate the volume in which the $\chi$ field particles are collected, using the birth rate of new phase bubbles and taking into account the probability of the birth of a new phase bubble in a remaining bubble of the old phase during a phase transition (i.e., estimate volume from which the cosmological Q-ball is born),  
\begin{equation}
    \label{valstar}
    V_{\star} = \xi \left(\frac{uA^{1/2}M^{*}_{pl}}{T^{2}_{c}L^{3/2}}\right)^{3}.
\end{equation}
It is worth noting that the charge of the cosmological Q-ball is determined by the number of charged particles of the $\chi$ field located in volume $V_{\star}$, therefore, it is possible to estimate the charge of the cosmological Q-ball:
\begin{equation}
\label{qstar}
    Q_{\star} = \eta_{\chi}\xi\frac{2\pi^2g_{*}}{45}\left(\frac{uA^{1/2}M_{pl}^{*}}{L^{3/2}T_{c}}\right)^{3},
\end{equation}
where $T_{c} \sim v$, $M_{Pl}^{*} = M_{Pl}/(1.66\sqrt{g_{*}})$, $g_{*} \sim 100$ is the effective number of degrees of freedom at the phase transition temperature, $u$ -- the velocity of the bubble walls of the new phase during the phase transition; $\xi\sim 1$, $L\sim 100$, $A\sim 1$ -- constants related to the calculation method (for more detailed calculations and physics, see Ref. \cite{Krylov:2013qe}). Thus, the charge of the cosmological Q-ball lies within
\begin{equation}
    \label{Q}
    Q_{min} < Q < Q_{\star},
\end{equation}
where $Q_{min}$ is defined by (\ref{Qmin}) and $Q_{\star}$ is defined by (\ref{qstar}).

\subsection{Distribution of Q-balls according to their charges}

In the previous section, it is assumed that all cosmological Q-balls are born with a charge of $Q_{\star}$. Now it is proposed to find the distribution of such Q-balls by their charges, taking into account (\ref{Q}). To do this, it is necessary to find the probability of 
the birth of a cosmological Q-ball with a charge greater than some $\bar{Q}$. Obviously, the probability of a new phase bubble appearing in a bubble of the old phase of volume $V$ is proportional to the volume of the considered area and the characteristic time of collapse of the bubble, since the longer the time and volume, the greater the chance of formation of a new phase bubble in the area \cite{Troitsky:2015mda},
\begin{equation}
    \label{probabilitynew}
    F_{b} = V\Gamma \frac{R}{u}.
\end{equation}

Since it is assumed that a cosmological Q-ball is born from a bubble of the old phase before a bubble of the new one appears in it, the following estimate can be given by
\begin{equation}
    \label{probability}
    V_{\star}\Gamma \frac{R_{\star}}{u} \sim 1.
\end{equation}

This estimate takes place, because when the maximum volume of the considered area of the old phase is reached, the probability of a bubble of a new phase appearing in it tends to unity. 

Next, it is necessary to find the probability of the birth of a Q-ball with a charge greater than $\bar{Q}$. From the definition of probability,
\begin{equation}
    \label{qprob}
    F = 1 - F_{b}.
\end{equation}
This expression can be divided by (\ref{probability})
\begin{equation}
    \label{qprobval}
    F = 1 - \frac{V\Gamma R/u}{V_{\star}\Gamma R_{\star}/u}.
\end{equation}

Let the particles of the field $\chi$ be evenly distributed throughout the early Universe, then $Q\sim V$, and therefore, (\ref{qprobval}) takes the form,
\begin{equation}
    \label{qprobq}
    F = 1 - \left(\frac{Q}{Q_{\star}}\right)^{4/3}  = \int \frac{dP}{dQ}dQ,
\end{equation}
where $dP/dQ$ is the probability of the birth of a Q-ball in the range from $Q$ to $Q+dQ$. To find this probability, it is necessary to consider the probability of the birth of a Q-ball with a charge greater than $\bar{Q}$ and with a charge of $\bar{Q}+dQ$ and find their difference,
\[
F(\bar{Q}+dQ) - F(\bar{Q}) = \int_{\bar{Q}+dQ}^{Q_{\star}} \frac{dP}{dQ}dQ - \int_{\bar{Q}}^{Q_{\star}} \frac{dP}{dQ}dQ.
\]

Hence, the distribution of cosmological Q-balls by their charges is

\begin{equation}
    \label{dPdQ}
    \frac{dP}{dQ} = -\frac{F(\bar{Q}+dQ)-F{\bar{Q}}}{dQ} = -\frac{dF}{dQ}.
\end{equation}
Thus,
\begin{equation}
    \label{nfromq}
    n(Q) \sim \alpha\int_{Q_{min}}^{Q}\frac{dP}{dQ}dQ, 
\end{equation}
where $\alpha$ is determined from the normalization condition,
\[
\alpha\int_{Q_{min}}^{Q_{\star}}Q\frac{dP}{dQ}dQ = Q_{\star}
\]
and is equal to 7/4. Here and further, it is assumed that $Q_{min}<<Q_{\star}$. Now it can be gotten the final expression for the distribution (\ref{nfromq}),
\begin{equation}
    \label{contribution}
    n(Q) \sim \frac{7}{4}\Big(\frac{Q}{Q_{\star}}\Big)^{4/3}.
\end{equation}
\section{ESTIMATE OF THE POTENTIAL PARAMETER}
\label{Estimation of the potential parameter}
In this section, the Lagrangian (\ref{FLSLagrangian}) parameter $v$ is estimated. As discussed above, Q-balls are dark matter, and therefore it is necessary to satisfy certain conditions.

Firstly, the mean cross section of the interaction of bulk dark matter should not exceed 1 cm$^{2}$/g \cite{Vogelsberger_2012, Rocha_2013, Zavala_2013, Elbert_2015, Troitsky:2015mda},
\begin{equation}
\label{sigmaavarege}
\langle\bar{\sigma}\rangle _{b} =\bar{\sigma}_{\star} \int^{1}_{0}\frac{x^{-1/4}x^{3/4}(1-x)}{x^{3/4}(1-x)}dx \approx 1.3\bar{\sigma}_{\star} \lesssim 1 \mbox{ cm$^{2}$/g},    
\end{equation}
where $x = Q/Q_{\star}$, $\bar{\sigma}_{\star} = \bar{\sigma}(Q_{\star})$. 

The mean cross section is 
\begin{equation}
    \label{sigma}
    \bar{\sigma}(Q) = \frac{\pi R^{2}_{Q}}{m_{Q}} = \frac{3}{8\sqrt{2}}v^{-3}Q^{-1/4}. 
\end{equation}

Then from (\ref{sigmaavarege}), taking into account (\ref{qstar}) and (\ref{sigma}), it can be gotten the lower limit for $v$,
\begin{equation}
    \label{vmin} 
    v_{min}\gtrsim \frac{10^{-7}u^{2/3}}{\eta_{\chi}^{1/9}}\mbox{ GeV}.
\end{equation}

Secondly, the present energy density of Q-balls should not exceed the present energy density of dark matter. The energy density of Q-balls obtained using (\ref{QballMass}) and (\ref{contribution}),
\begin{equation}
    \label{rhofromq}
    \rho = \int_{0}^{Q_{\star}}m_{Q}dn(Q) \sim Q^{25/12}.
\end{equation}
This function has a maximum at the upper limit of integration $Q_{\star}$, therefore, for a rough estimate, it is assumed that cosmological Q-balls with $Q_{\star}$ charge dominate in dark matter.

\begin{figure}[htb]
    \centering
    \includegraphics[width=1\columnwidth]{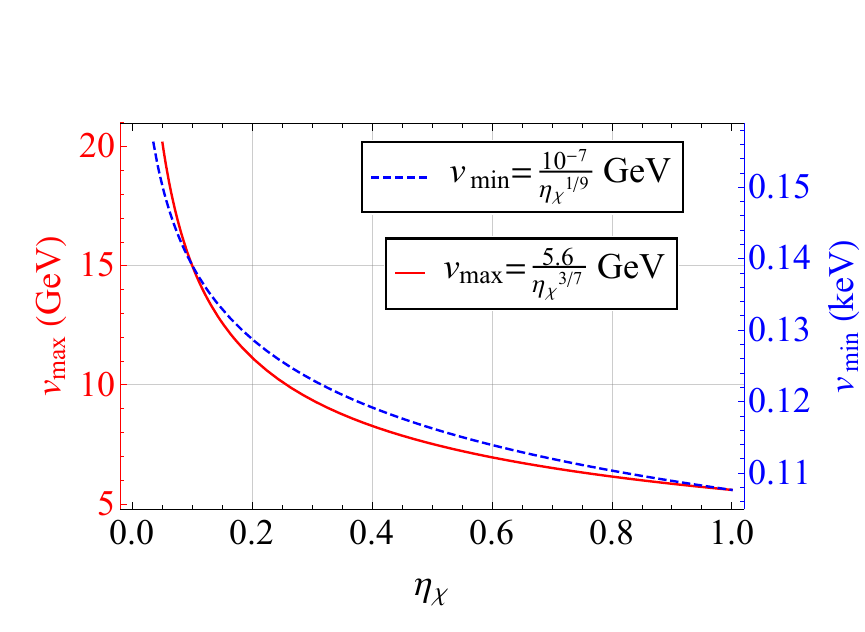}
    \caption{Dependence of cosmological constraints on the parameter $v$ of the Lagrangian (\ref{FLSLagrangian}) on the magnitude of the asymmetry of the particles of the field $\chi$ $\eta_{\chi}$ in the case of the velocity of cosmological Q-balls $u=1$ and $T_{c} =v$. $v_{min}$ is the lower limit and is represented in keV (blue dashed line), $v_{max}$ is the upper limit and is represented in GeV (red line).}
    \label{vgraph}
\end{figure}

There is a connection between the energy density of Q-balls and the cross section, given in \cite{Troitsky:2015mda}. Using this ratio,  the upper limit for $v$ can be found
\begin{equation}
\label{rho}
    \rho_{DM} \gtrsim \frac{4\sqrt{2}\pi}{3}v\cdot Q_{\star}^{-1/4}\eta_{\chi}s_{0},
\end{equation}
where $\rho_{DM} = 10^{-6}\mbox{ GeV/cm}^{3}$ is the modern density of dark matter, $s_{0} = 3\times 10^{3}$ cm$^{-3}$ is the modern density of entropy. Thus, the upper limit on $v$ is
\begin{equation}
    \label{vmax}
    v_{max} \lesssim \frac{5.6u^{3/7}}{\eta_{\chi}^{3/7}}\mbox{ GeV}.
\end{equation}
Dependence of the maximum and minimum values of $v$ on the asymmetry is shown in the Fig. \ref{vgraph}.
In some works it is proposed to estimate the asymmetry of the particles of the field $\chi$ as a quark asymmetry $3\Delta_{B}\approx 2.7\times 10^{-10}$, and the velocity of cosmological Q-balls as $u\sim 1$. In the case of such estimate, $v$ is \cite{Krylov:2013qe}
\begin{equation}
    \label{quarkasv}
    1.2\mbox{ keV} \lesssim v \lesssim 70.4 \mbox{ TeV}.
\end{equation}
Now, accordingly, within the framework of this estimate, it is possible to find restrictions on the mass of cosmological Q-balls using (\ref{QballMass}), (\ref{qstar}), and (\ref{quarkasv})
\begin{equation}
    \label{mstar}
    3\cdot10^{-5}\mbox{ kg }\lesssim m_{\star} \lesssim 8\cdot10^{8} \mbox{ kg.}
\end{equation}
\section{Q-BALLS MERGING}
\label{Q-BALLS MERGING}
It follows from (\ref{mstar}) that cosmological Q-balls cannot claim to be exotic objects of stellar mass; therefore, it is necessary to consider their inelastic interaction, which occurs with a probability of $\sim 50\%$ \cite{Multamaki:2000ey}, \cite{Multamaki:2000qb}.
\subsection{Flat expanding Universe}
\label{Flat non-expanding Universe}
In this model, it is assumed that after the phase transition shown in Fig. \ref{combuction}, the early flat expanding Universe is uniformly filled with cosmological Q-balls with the same charges (\ref{qstar}) and the same velocities $u$. It is proposed to select one cosmological Q-ball (hereinafter referred to as the selected Q-ball), which absorbs other cosmological Q-balls during the evolution of the Universe, and based on this simple mechanism, obtain an estimate of the mass of the selected Q-ball.

At the initial stage, the velocity of the cosmological Q-ball (it is also the velocity of the wall of the region from which the cosmological Q-ball is born during the phase transition) $u$, the asymmetry $\eta_{\chi}$ of the particles of the field $\chi$, the Lagrangian parameter $v$, and the phase transition temperature $T_{c}\sim v$ are considered free parameters limited only by simple physical considerations. Thus, the velocity of the cosmological Q-ball $u$ should not exceed the speed of light
\[
0 < u < 1;
\]
the asymmetry of the $\chi$ field is limited as follows from the definition of asymmetry (\ref{asymmetry}):
\[
0 \lesssim \eta_{\chi} \lesssim 1. 
\]
The parameter $v$, in addition to the restrictions shown in Fig. \ref{vgraph}, is limited from above by the temperature of the Planck epoch, and from below, by the temperature of the present Universe. This is also true for the phase transition temperature $T_{c}$ \cite{Fixsen_2009}, 
\[
10^{18 \div 19} \mbox{ GeV} \gtrsim v\sim T_{c} \gtrsim 2.3 \times 10^{-13}\mbox{ GeV} \approx 2.7\mbox{ K}.
\]
Firstly, the model assumes that the early flat expanding Universe is uniformly filled with cosmological Q-balls, which are born in  bubbles of a new phase of the volume $V_{\star}$, which, in turn, is determined by (\ref{valstar}). Due to this fact, at the initial stage, there is on average one cosmological Q-ball in this volume. Secondly, due to the evolution of the Universe, the concentration of Q-balls is not a constant value (here, the concentration change from the interaction of the selected Q-ball with cosmological Q-balls is considered insignificant). To account for changes in concentration due to the expansion of the Universe, it is necessary to introduce a scale factor, which is determined from the Friedman equation and is equal to 
\begin{equation}
    \label{FridmanSolution}
    a = a_{0}\Big(\frac{\Omega_{M}}{\Omega_{\Lambda}}\Big)^{1/3}(\sinh[\frac{3}{2}\sqrt{\Omega_{\Lambda}}H_{0}t])^{2/3},
\end{equation}
where $\Omega_{M}=0.24$, $\Omega_{\Lambda}=0.76$, $H_{0}=2.2\times 10^{-18} \mbox{ s}^{-1}$  \cite{rubakov2011introduction}, \cite{Planck2020}. Normalization of the scale factor
$a_{0}$ is determined from the condition,
\[
a(t_{0})=1,
\]
where $t_{0} = 14$ Gyr is the present epoch of the Universe. Then, we can find the concentration of cosmological Q-balls in the Universe,
\begin{equation}
    \label{nexp}
    n_{\star} = \frac{1}{V_{\star}a^{3}(t)}.
\end{equation}

It is assumed that cosmological Q-balls with velocities of the order of $u$ merge with a selected Q-ball; therefore, their interaction is determined by the geometric cross section of the selected Q-ball, which is defined as follows:
\begin{equation}
\label{sigmageom}
    \sigma = \frac{\pi R_{Q}^{2}(t)}{2},
\end{equation}
where $R_{Q}(t)$ is determined from (\ref{QballRadius}) and obviously grows over time due to the absorption of the selected Q-ball of cosmological Q-balls. The probability of an inelastic interaction of bulk dark matter is taken into account here, equal to $\sim 50 \%$ \cite{Multamaki:2000ey}, \cite{Multamaki:2000qb}.

Now an equation describing change of the charge  (and, accordingly, mass) during the interaction of the selected Q-ball with cosmological Q-balls can be made.  Let the selected cosmological Q-ball absorb charge $\Delta Q$ in time $\Delta t$. Obviously, the greater the concentration $n$ of cosmological Q-balls in a given region and the larger the geometric cross-section of the interaction $\sigma$, the selected Q-ball absorbs the greater charge 
\begin{equation}
\label{deltaQ}
    \Delta Q = Q_{\star}\sigma n u\Delta t,
\end{equation}
 Now it is necessary to make the limit transition,
\[
\Delta Q \to 0, \:\: \Delta t \to 0.
\]
Thus, the law of change of charge of the selected Q-ball due to its absorption of cosmological Q-balls in time is
\begin{equation}
    \label{QballsmergingExp}
    \dot{Q}= Q_{\star}u\sigma(Q)n_{\star},
\end{equation}
where $u$ is the velocity of the cosmological Q-ball, equal to the velocity of the walls during the first order phase transition and collapse to a cosmological Q-ball, $Q_{\star}$ is a charge of cosmological Q-ball determined by (\ref{qstar}), $n_{\star}$ is the concentration determined by (\ref{nexp}), and $\sigma(Q)$ is the geometric cross section of the interaction of Q-balls determined by (\ref{sigmageom}).
 For the correct formulation of the mathematical problem, it is necessary to add one initial condition, which is determined from a simple physical consideration: at the time when the phase transition occurred $t_{c}\sim 0$ (hereafter, for simplicity, it is assumed that the phase transition occurs in the early Universe), the charge of the selected Q-ball is
\begin{equation}
    \label{Qfromzero}
    Q(t_{c}) = Q_{\star}.
\end{equation}
Combining (\ref{QballsmergingExp}) and (\ref{Qfromzero}), a correct mathematical problem can be obtained 
\begin{equation}
    \begin{cases}
    \label{merging}
            \dot{Q}= Q_{\star}u\sigma(Q)n_{\star},
            \\
             Q(t_{c}) = Q_{\star}.
    \end{cases}
\end{equation}
Using (\ref{QballRadius}), (\ref{FridmanSolution}), (\ref{nexp}) and (\ref{sigmageom}) and separating the variables it can be obtained from (\ref{merging})
\begin{equation}
    \label{ExpSolFull}
    \int\frac{dQ}{Q^{1/2}}=\frac{\pi Q_{\star}u\Omega_{\Lambda}}{4v^{2}V_{\star}a_{0}^{3}\Omega_{M}}\int\sinh{\left[\frac{3}{2}\sqrt{\Omega_{\Lambda}}H_{0}t\right]dt},
\end{equation}
and the solution of (\ref{merging}) after integration (\ref{ExpSolFull}) is 
\begin{equation}
    \label{ExpSol}
Q(t)=\left(- \frac{\pi Q_{\star} u \sqrt{\Omega_{\Lambda}}}{12v^{2}V_{\star}a_{0}^{3}\Omega_{M}H_{0}}\coth{\left[\frac{3}{2}\sqrt{\Omega_{\Lambda}}H_{0}t\right]}+C\right)^{2},
\end{equation}
where $C$ is the integration constant that can be found from the initial condition  (\ref{Qfromzero}) and equal to
\begin{equation}
C=\frac{\pi Q_{\star}u\sqrt{\Omega_{\Lambda}}}{12v^{2}V_{\star}a_{0}^{3}\Omega_{M}H_{0}}\coth{\left[\frac{3}{2}\sqrt{\Omega_{\Lambda}}H_{0}t_{c}\right]}+\sqrt{Q_{\star}},
\end{equation}
where $t_{c}$ is the time of the phase transition, $V_{\star}$ is defined by  (\ref{valstar}) and $Q_{\star}$ is defined by (\ref{qstar}).

In any physically meaningful configuration of the free parameters $v$, $u$, $T_{c}$, and $\eta_{\chi}$ the solution (\ref{ExpSol}) is almost time independent, and therefore, taking into account (\ref{QballMass}), it is impossible to obtain the selected Q-ball of stellar mass, and free cosmological Q-balls in the case of a flat expanding Universe almost do not interact.
\subsection{Galaxies}
As it is shown in the previous section, free cosmological Q-balls almost do not interact with the selected Q-ball, and, as a result, the selected Q-ball almost does not change its mass during the evolution of the Universe. In fact, this is primarily due to the presence of a scale factor in (\ref{nexp}), which strongly suppresses the interaction of cosmological Q-balls with a selected Q-ball.

It is known that the expansion of the Universe does not affect gravitationally bound structures in the Universe, for example, galaxies. However, it is known that galaxies are formed primarily from dark matter \cite{Silk}. Let all dark matter be represented by cosmological Q-balls after the first order phase transition. In this case, it can be assumed that after the phase transition, cosmological Q-balls participate in the formation of galaxies and, therefore, are in the gravitational potential of galaxies (the mechanisms of galaxy formation and cosmological Q-balls entering the gravitational potential of galaxies are not discussed in detail in this work) \cite{Conselice_2014}. This model assumption allows us to get rid of the scale factor in (\ref{nexp}) and, as a result, locally get rid of the suppression of the Q-balls interaction. Further, being in the gravitational potential of the galaxy, Q-balls of different charges begin to be absorbed by the selected Q-ball, which, in turn, is expected to acquire a mass of the order of the mass of the Sun, so that such Q-balls can fill the mass gap between light black holes and massive neutron stars.

Now it is necessary to discuss in detail the formulation of the law of interaction of Q-balls of different charges with a selected Q-ball within the framework of a model of their interaction in the gravitational potential of galaxies, which is constructed by analogy with the law (\ref{merging}). Firstly, it is necessary  to change the concentration (\ref{nexp}). In addition to the absence of a scale factor, the volume in which one Q-ball is located also changes. This is due to the fact that dark matter in galaxies is unevenly distributed. In this work, it is assumed that Q-balls of dark matter obey the Burkert profile to avoid the presence of a singularity in the dark matter density at the galaxy center. This profile is
\begin{equation}
\label{NFWprofile}
    \rho(r) = \frac{\rho_{b}}{\left(1+\frac{r}{R_{s}}\right)\left(1+\frac{r^{2}}{R_{s}^{2}}\right)},
\end{equation}
where $R_{s} \approx 0.3$ kpc -- radius of galaxy core, $r$ is the distance from the center of the galaxy and $\rho_{b} = 2.34 \times10^{4}$ GeV/$\mbox{cm}^{3}$ is found from the condition that total mass of dark matter halo is $\sim 10^{12} \mbox{ M}_{\odot}$ (hereafter, the characteristic parameters for the Milky Way are used, since this is the most studied galaxy) \cite{Burkert_1995}, \cite{refId0}.
\begin{equation}
    \label{rhob}
    \int^{200 \mbox{ kpc}}_{0  \mbox{ kpc}} 4 \pi r^{2} \rho(r)dr \sim 10^{12} \mbox{ M}_{\odot},
\end{equation}
where the upper limit of integration is equal to the radius of the dark matter halo.
Then, the concentration is
\begin{equation}
    \label{concentration}
    n(r) = \frac{\rho(r)}{m_{Q}(Q)},
\end{equation}
where the concentration change at Q-balls merger is taken into account due to the charge (and hence, time) dependence of the Q-ball masses $m_{Q}(Q)$ which are given by (\ref{QballMass}).

Secondly, Q-balls are located in the gravitational field of the galaxy and represent cold dark matter. This imposes some restrictions on the velocity of the Q-balls (but not on the velocity of the walls during the phase transition!). The characteristic orbital velocity of stars in the Milky Way $u_{\star} \approx 220$ km/s. A natural assumption is made here that Q-balls in the gravitational potential of the galaxy move at velocities characteristic of stars in this potential.

Thirdly, due to the fact that the phase transition occurs due to the cooling of the Universe, additional restrictions appear on the Lagrangian parameter $v$ and the temperature of the phase transition $T_{c}$, due to the fact that the phase transition must occur before the formation of the first galaxies. Since the first galaxies appear at time $t\sim 100\mbox{ Myr}\div1$ Gyr, the following restriction can be set
\begin{equation}:
    \label{tcconstraint}
    10^{-12}\mbox{ GeV} \lesssim v\sim T_{c} \lesssim 10^{18 \div 19} \mbox{ GeV},
\end{equation}
where the lower limit corresponds to the temperature of the Universe at time $t\sim 1$ Gyr \cite{rubakov2011introduction}.

Now a mathematical problem can be started setting in the same way as in (\ref{merging}). The cross section does not depend on model assumptions; therefore, it has the form (\ref{sigmageom}). The characteristic velocity of Q-balls in a galaxy is equal, as discussed above, to the characteristic orbital velocity of stars in the gravitational potential of galaxies $u_{\star} \approx 220$ km/s. The concentration is given by the expression (\ref{concentration}), taking into account the density distribution of dark matter in the galaxy (\ref{NFWprofile}) and time dependence of concentration due to Q-balls merging. Due to the fact that Q-balls are not absorbed by the selected Q-ball before they enter the gravitational potential of the galaxy, it is convenient to solve the equation within $t\in [0; 13]$ Gyr, where it is taken into account that the first galaxies appear at time 1 Gyr. Thus, a differential equation similar to (\ref{merging}) is
\begin{equation}
    \label{mergingNFWfull}
    \begin{cases}
                \dot{Q}= \sum\limits_{k} kQ_{\star}u_{\star}\sigma(kQ_{\star})n(kQ_{\star}), \:\: t \in [0; 13] \mbox{ Gyr}, \: k \in \mathbb{N},
                \\
             Q(0) = Q_{\star},
             \end{cases}
\end{equation}
where it is taken into account that the selected Q-ball can absorb Q-balls with different charges $kQ{\star} \leq Q(t)$. Let us consider the behavior of the $k$th term of (\ref{mergingNFWfull}). Using (\ref{QballRadius}), (\ref{QballMass}), (\ref{sigmageom}), and (\ref{concentration}), it can be obtained 
\begin{equation}
    \label{kterm}
    \dot{Q} \sim (kQ_{\star})^{3/4}.
\end{equation}
From the view of (\ref{kterm}), it can be seen that the main contribution to the charge changing in (\ref{mergingNFWfull}) is made by the term with maximal $k$ which corresponds to the situation when $kQ_{\star} = Q(t)$. According to this, (\ref{mergingNFWfull}) can be simplified for estimates,
\begin{equation}
    \label{mergingNFW}
        \begin{cases}
                \dot{Q}= Qu_{\star}\sigma(Q)n(Q), \:\: t \in [0; 13] \mbox{ Gyr},
                \\
             Q(0) = Q_{\star}.
             \end{cases}
\end{equation}
Using (\ref{QballRadius}), (\ref{QballMass}), (\ref{sigmageom}), and (\ref{concentration}) and separating the variables it can be obtained from (\ref{mergingNFW})
\begin{equation}
    \label{solutionNFWfull}
    \int \frac{dQ}{Q^{3/4}} = \int\frac{3u_{\star}\rho(r)}{16\sqrt{2}v^{3}}dt,
\end{equation}
and the solution of (\ref{mergingNFW}) after integration (\ref{solutionNFWfull}) and taking into account the initial condition from (\ref{mergingNFW}) is 
\begin{equation}
    \label{solutionNFW}
    Q(t,r) = \left(\frac{3u_{\star}\rho(r)}{64\sqrt{2}v^{3}}t+Q_{\star}^{1/4}\right)^{4},
\end{equation}
where $Q_{\star}$ is defined by (\ref{qstar}), $\rho(r)$ is defined by (\ref{NFWprofile}).
\section{SOLUTION ANALYSIS}
\label{SOLUTION ANALYSIS}
In this section, the solution (\ref{solutionNFW}) obtained within the framework of the model of merging Q-balls with a selected Q-ball in the gravitational potential of the galaxy is analyzed.

To obtain the mass of the selected Q-ball, the solution (\ref{solutionNFW}) must be substituted into the expression for the mass (\ref{QballMass}). Thus, the final expression for the mass is
\begin{equation}
    \label{massresult}
    \begin{split}
    m_{Q}(v,u,\eta_{\chi},u_{\star},T_{c},r, t) 
    = 
    \\
    = \frac{4\sqrt{2}\pi}{3}v\left(\frac{3u_{\star}\rho(r)}{64\sqrt{2}v^{3}}t+Q_{\star}^{1/4}\right)^{3},
    \end{split}
\end{equation}
where is $Q_{\star} = Q_{\star}(v,u,\eta_{\chi},T_{c}) $ is defined by (\ref{qstar}), and $\rho =\rho(r)$ from (\ref{NFWprofile}). The main motivation of this work is to answer the question ''Can dark-matter Q-balls grow to the mass gap masses?''; therefore, it is necessary to satisfy the following limitation:
\begin{equation}
    \label{mconstraint}
    m_{Q}(v,u,\eta_{\chi},u_{\star},T_{c},r)\bigg|_{t =13\mbox{ Gyr}}  \gtrsim 1 M_{\odot}.
\end{equation}
As a first approximation, there are many sets of free parameters that satisfy this restriction; however, the work presents several restrictions on free parameters, in particular (\ref{sigmaavarege}), (\ref{rho}), and (\ref{tcconstraint}), which must be satisfied. In this case, the set of free parameters is narrowed, but, nevertheless, different suitable solutions still exist. As an example of one of these sets of free parameters, the following values are presented, which are referred to in this work as the most successful set of free parameters and which are used in further calculations:
\begin{equation}
\label{parameters}
    \begin{cases}
        v \approx 10^{-7} \mbox{ GeV}, \\
        u = 1, \\
        \eta_{\chi} = 1, \\
        u_{\star} = 0.0007, \\
        T_{c} \approx 10^{-7} \mbox{ GeV}. \\
    \end{cases}
\end{equation}

It can be seen that the requirement of $v \sim T_{c}$ is well satisfied, and the mean cross section from (\ref{sigma}) with this set of parameters is equal $0.8 \mbox{ cm}^{2}/\mbox{g}$ and hence, also satisfies restriction (\ref{sigmaavarege}). Forging of Q-balls in this scenario occurs at the time $t_{c}= 10\mbox{ s}\div 3\times 10^{5}$ years since the big bang (this corresponds to the photon epoch) \cite{rubakov2011introduction}.

The mass limit (\ref{mconstraint}) is satisfied. From the form of (\ref{massresult}) taking into account (\ref{NFWprofile}) and (\ref{parameters}), it can be understood that the further away the selected Q-ball is from the center of the galaxy, the smaller its mass. For example, in the case of distance $r = 0.05$ kpc from the galaxy center, the present mass of Q-balls is
\begin{equation}
    m_{Q}\bigg|_{r = 0.05 \mbox{ kpc}} \approx 5 \mbox{ M}_{\odot}.
\end{equation}

It can be obtained by using (\ref{QballRadius}) and (\ref{solutionNFW}) that radius of selected Q-ball is

\begin{equation}
    \label{radiusresult}
        \begin{split}
    R_{Q}(v,u,\eta_{\chi},u_{\star},T_{c},r,t) 
    = 
    \\
    = \frac{1}{\sqrt{2}v}\left(\frac{3u_{\star}\rho(r)}{64\sqrt{2}v^{3}}t+Q_{\star}^{1/4}\right),
    \end{split}
\end{equation}
where is $Q_{\star} = Q_{\star}(v,u,\eta_{\chi},T_{c}) $ is defined by (\ref{qstar}), and $\rho =\rho(r)$ from (\ref{NFWprofile}). Hence, the present radius of Q-balls at distance $r = 0.05$ kpc from the galaxy center is
\begin{equation}
    R_{Q}\bigg|_{r = 0.05 \mbox{ kpc}} \sim 10^{9}\mbox{ km}
\end{equation}
which is roughly equal to the distance from the Sun to Neptune.

\begin{figure}[htb]
    \centering
    \includegraphics[width=\columnwidth]{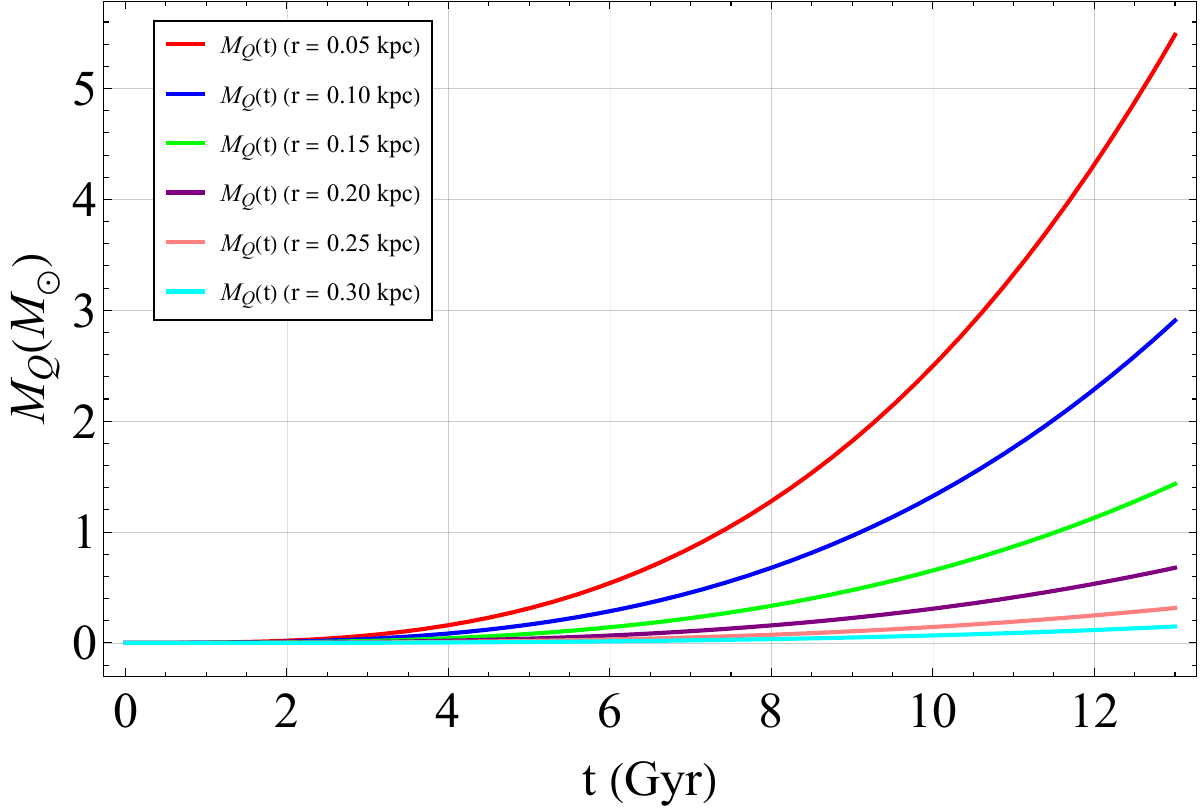}
    \caption{The evolution of selected Q-ball mass (\ref{massresult}) depending on the distance from the center of the galaxy $r$ and time $t$. The red curve corresponds to the situation when the selected Q-ball is located at a distance of $0.05$ kpc. The blue curve corresponds to the situation when the selected Q-ball is located at a distance of $0.10$ kpc. The green curve corresponds to the situation when the selected Q-ball is located at a distance of $0.15$ kpc. The purple curve corresponds to the situation when the selected Q-ball is located at a distance of $0.20$ kpc. The pink curve corresponds to the situation when the selected Q-ball is located at a distance of $0.25$ kpc. The cyan curve corresponds to the situation when the selected Q-ball is located at a distance of $0.30$ kpc.}
    \label{massevolutionpic}
\end{figure}

\begin{figure}[htb]
    \centering
    \includegraphics[width=\columnwidth]{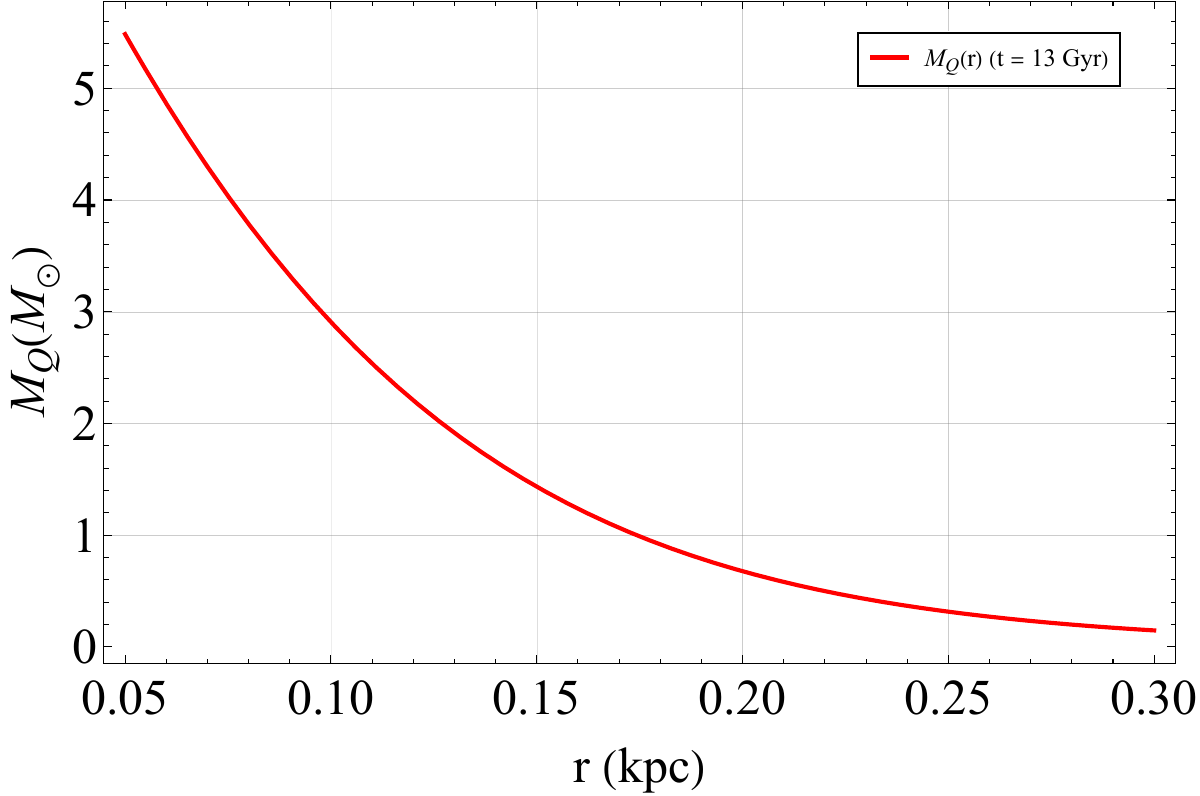}
    \caption{The mass profile of the selected Q-ball (\ref{massresult}) as a function of the distance from the center of the galaxy $r$ at time $t = 13$ Gyr (red curve).}
    \label{massfromrpic}
\end{figure}

\begin{figure}[htb]
    \centering
    \includegraphics[width=\columnwidth]{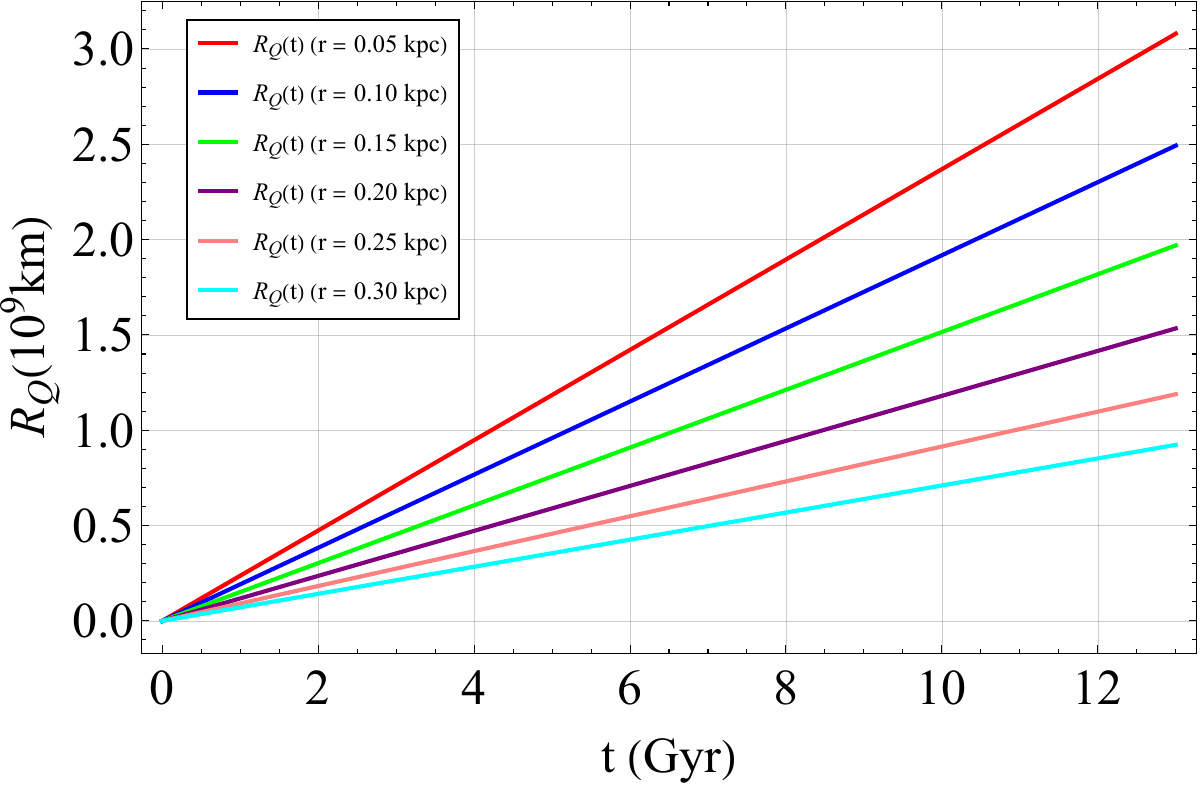}
    \caption{The evolution of selected Q-ball radius (\ref{radiusresult}) depending on the distance from the center of the galaxy $r$ and time $t$. The red curve corresponds to the situation when the selected Q-ball is located at a distance of $0.05$ kpc. The blue curve corresponds to the situation when the selected Q-ball is located at a distance of $0.10$ kpc. The green curve corresponds to the situation when the selected Q-ball is located at a distance of $0.15$ kpc. The purple curve corresponds to the situation when the selected Q-ball is located at a distance of $0.20$ kpc. The pink curve corresponds to the situation when the selected Q-ball is located at a distance of $0.25$ kpc. The cyan curve corresponds to the situation when the selected Q-ball is located at a distance of $0.30$ kpc.}
    \label{radiusevolutionpic}
\end{figure}

\begin{figure}[htb]
    \centering
    \includegraphics[width=\columnwidth]{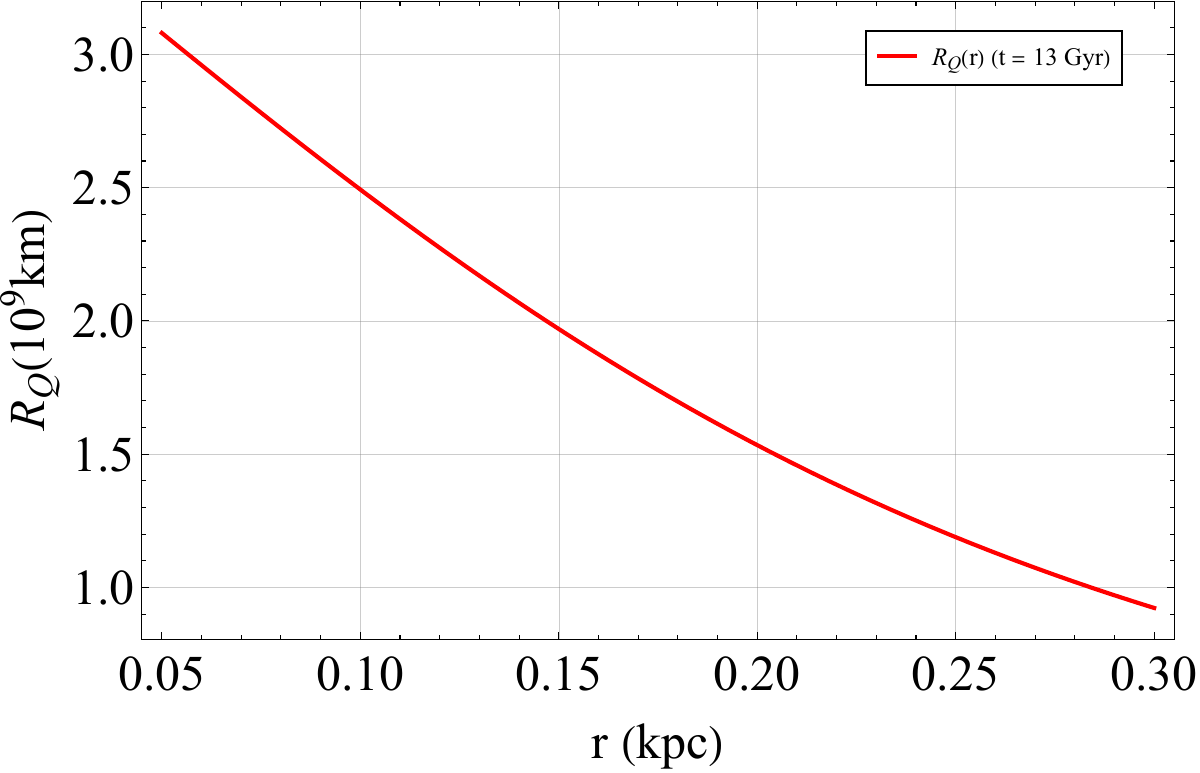}
    \caption{The radius profile of the selected Q-ball (\ref{radiusresult}) as a function of the distance from the center of the galaxy $r$ at time $t = 13$ Gyr (red curve).}
    \label{Rfromronpic}
\end{figure}

The Fig. \ref{massevolutionpic} demonstrates the evolution of mass of selected Q-ball as a function of time $t$ and distance from the center of the galaxy $r$. The Fig. \ref{massfromrpic} demonstrates the mass profile of the selected Q-ball as a function of the distance from the center of the galaxy $r$ at the time $t = 13$ Gyr. The Fig. \ref{radiusevolutionpic} demonstrates the evolution of radius of selected Q-ball as a function of time $t$ and distance from the center of the galaxy $r$. The Fig. \ref{Rfromronpic} demonstrates the radius profile of the selected Q-ball as a function of the distance from the center of the galaxy $r$ at the time $t = 13$ Gyr. It can be seen that the main interaction of the selected Q-ball with other Q-balls occurs in the central part of the galaxy.

It can be seen from form of (\ref{massresult}) with substitution (\ref{NFWprofile}) and from Fig. \ref{massfromrpic} that selected Q-ball mass decreases with distance from galaxy center $r$, so it can be obtained the distance from galaxy center where present selected Q-ball mass with parameters (\ref{parameters}) becomes less then $1 \mbox{ M}_{\odot}$,
\begin{equation}
    \label{stellarR}
    m_{Q}(r) = 1 \mbox{ M}_{\odot} \Rightarrow r \approx 0.17 \mbox{ kpc}.
\end{equation}
Hence, the present total number of stellar-mass Q-balls is
\begin{equation}
     \label{Nqstellar}
     N_{Q}^{stellar} = \int_{0 \mbox{ kpc}}^{0.17 \mbox{ kpc}} \frac{4 \pi r^{2} \rho(r)}{m_{Q}(r)}dr \approx 4 \times10^{9},
 \end{equation}
where the upper limit of integration is determined from (\ref{stellarR}).

In the approximation that the mass of Q-balls is completely absorbed by selected Q-ball during the merger, the mass of the selected Q-ball with parameters (\ref{parameters}) cannot be less than two masses of cosmological Q-balls with the same set of free parameters after the merger (it corresponds to merger of two cosmological Q-balls). According to this the distance from the center of the galaxy at which the merger of Q-balls stops can be found from equation,
\begin{equation}
    \label{nointeractionsR}
        m_{Q}(r) = 2m_{\star} \Rightarrow r \approx 16 \mbox{ kpc},
\end{equation}
where $m_{\star} = m_{Q}(Q_{\star})$. Hence, the present total number of all Q-balls in galaxy is
   \begin{equation}
     \label{Nqtotal}
     \begin{split}
     N_{Q}^{total} = \int_{0 \mbox{ kpc}}^{16 \mbox{ kpc}} \frac{4 \pi r^{2} \rho(r)}{m_{Q}(r)}dr +  
     \\
     +\int_{16 \mbox{ kpc}}^{200 \mbox{ kpc}} \frac{4 \pi r^{2} \rho(r)}{m_{\star}}dr \sim  10^{24},
     \end{split}
 \end{equation}
 where the upper limit of integration in the second term is determined from the radius of the dark matter halo, and the upper limit of integration in the first term and the lower in the second is determined from (\ref{nointeractionsR}).
 
Thus, the present masses of Q-balls lie within
  \begin{equation}
     \label{masses}
     m_{\star} \approx 10^{-13} \mbox{ M}_{\odot} \leq m_{Q}(r) \lesssim  m_{Q}(r=0) \approx10 \mbox{ M}_{\odot}, 
 \end{equation}
and the present radii of Q-balls lie within
 \begin{equation}
     \label{radii}
     R_{Q}(Q_{\star}) \approx 9\times10^{4}\mbox{ km} \leq R_{Q}(r) \lesssim R_{Q}(r=0) \approx 4\times10^{9}\mbox{ km.}
 \end{equation}

 The main parameters of Q-balls in the present epoch are presented in the Table \ref{table}.
\begin{table*}[!ht]
\caption{The main parameters of Q-balls in the modern era.}
\label{table}
\begin{ruledtabular}
\begin{tabular*}{\textwidth}{@{\extracolsep{\fill}}lllllll@{}}
The most successful set of free parameters: & \multicolumn{1}{c}{$v\sim 10^{-7}$ GeV} & \multicolumn{1}{c}{$T_{c} \sim 10^{-7}$ GeV} & \multicolumn{1}{c}{$\eta_{\chi} = 1$} & \multicolumn{1}{c}{$u = 1$} & \multicolumn{1}{c}{$u_{\star} = 0.0007$} & $m_{\star} \sim 10^{-13} \mbox{ M}_{\odot}$   \\
\hline
  $r$(kpc) & 0.05 & 0.17 & 0.30 & 8 & 16 & 200
        \\
        $M_{Q}(\mbox{M}_{\odot})$  & 5 & 1 & 0.15 & $10^{-12}$ & $10^{-13}$ & $10^{-13}$
        \\
        $R_{Q}$(km) & $ 10^{9}$ & $10^{9}$ & $10^{9}$ & $10^{5}$ & $10^{5}$ & $10^{5}$
        \\
\end{tabular*}
\end{ruledtabular}
\end{table*}
\section{SUMMARY AND DISCUSSIONS}
It is worth noting that it could be assumed that cosmological Q-balls are born during a phase transition immediately with masses of the order of one mass of the Sun. In this case, the need to compile the law of interaction of cosmological Q-balls with a selected Q-ball (\ref{merging}) has disappeared, but any physically meaningful configuration of the free parameters $v$, $u$, $T_{c}$ and $\eta_{\chi}$ prohibits this scenario so cosmological Q-balls cannot claim to be objects of stellar mass; therefore, it is necessary to consider their interaction.

The simplest case assumes that some selected cosmological Q-ball absorbs other cosmological Q-balls in the flat expanding Universe. The trivial equation of such interaction (\ref{QballsmergingExp}) shows that for any physically meaningful set of free parameters that include the Lagrangian (\ref{FLSLagrangian}) parameter $v$, the asymmetry $\eta_{\chi}$ of the particles of the field $\chi$, the velocity of the walls of the bubble of the new phase $u$ (and, accordingly, the velocity of free cosmological Q-balls), the temperature of the phase transition $T_{c}$, the selected cosmological Q-ball is unable to gain any significant mass. First of all, this is due to the expansion of the Universe, due to which the ''gas'' of cosmological Q-balls becomes extremely sparse. Thus, within the framework of such a trivial approximation, it is impossible that cosmological Q-balls gain stellar masses and explain the unusual gravitational wave events received by LIGO and Virgo.

Due to the fact that the expansion of the Universe suppresses the interaction of cosmological Q-balls, it is proposed to consider the merger of cosmological Q-balls in gravitationally bound objects -- galaxies. This makes it possible to neglect the expansion of the Universe, but adds new restrictions, for example, on the epoch of the phase transition, the concentration of cosmological Q-balls and their velocity. A simple equation (\ref{mergingNFW}) of Q-balls interaction in the galaxy gravitational potential is compiled, the solution (\ref{solutionNFW}) of which leads to the following results: Q-balls are able to gain the necessary masses of the order of the mass of the Sun, however, with any physically meaningful set of free parameters, such Q-balls have radii at present epoch of the order of the radius of the Solar system. Such a value of the present radii of Q-balls are significantly larger than the Schwarzschild radius for one mass of the Sun, which leads to the fact that Q-balls turn out to be very ''loose'', and configuration of present Q-balls is similar to clouds of dark matter, so unusual gravitational wave events received by LIGO and Virgo cannot be explained in the framework of presented model.  Estimates of the present populations of such Q-balls in the dark matter halo (\ref{Nqstellar}) and (\ref{Nqtotal}) are also obtained. Given that the present mass of the selected Q-ball decreases rapidly with distance to the center of the galaxy (Fig. \ref{massfromrpic}), the present population of Q-balls can be roughly divided into two groups: Q-balls of stellar masses, which are located at a distance of up to 0.17 kpc from the center of the galaxy and the number of which is estimated as (\ref{Nqstellar}) and Q-balls of small masses (of the order of the mass of the cosmological Q-ball $m_{\star}$ at distances of the order of 10 kpc), which are located at distances greater than 0.17 kpc. It is worth noting that these present populations, taking into account (\ref{masses}), are consistent with the results of observing microlensing events \cite{Mroz:2024mse}, as a result of which they may be of potential interest in different sections of astrophysics and cosmology.

It is worth discussing some inaccuracies of presented models. Firstly, estimate of the volume from which the cosmological Q-ball are born (\ref{valstar}) and its charge (\ref{qstar}) are inaccurate. In addition to the inaccuracies associated with calculations, which are described in detail in Ref. \cite{Krylov:2013qe}, certain inaccuracies are carried by the parameter $g_{*}$, which is responsible for the effective number of degrees of freedom and which depends on temperature. Clarifying this parameter is a separate work; therefore, it is decided to use the value proposed in \cite{Krylov:2013qe} for estimates. Secondly, the Milky Way is chosen as a typical galaxy, as it is the most studied. In fact, there is a vast population of different types of galaxies that may differ significantly from ours in their configuration. This leads to the fact that the concentration of Q-balls in galaxies (\ref{concentration}) can differ significantly depending on the galaxy, which may affect the final result. Despite these inaccuracies, the results of this work show that the scenario in which Q-balls acquire stellar masses at present time is not forbidden.

In addition to the analysis of the considered models, other important results are also obtained in this work. The distribution of cosmological Q-balls by their charges (\ref{contribution}), presented in \cite{Troitsky:2015mda}, is clarified. Also, a restriction is obtained on the parameter $v$ of the Lagrangian (\ref{FLSLagrangian}), which is shown in the Fig. \ref{vgraph}. 

Thus, within the presented models, Q-balls may have masses around one solar mass with radii around the radius of the Solar system at present epoch, but this does not mean that dark matter Q-balls cannot explain the unusual gravitational wave signals obtained by LIGO and Virgo. Only one simple model of Friedberg-Lee-Searlen Q-balls is considered in this work; however, there are other models that admit the existence of nontopological solitons, which can give a different result even within the framework of the proposed merger models \cite{q-balls}, \cite{Coleman:1985ki}, \cite{Heeck_2023}. Also, this study may help to narrow the scope of the search for solutions to the problems of black hole mass gap and unusual gravitational wave signals and add new limitations for future research.
\begin{acknowledgments}
The Author is grateful to Sergey Troitsky for the original idea and for fruitful discussions on the manuscript. This work was supported by the Russian Science Foundation (RSF) Grant No. 22-12-00215.
\end{acknowledgments}
\bibliography{Qballs}
\end{document}